\documentclass[a4,11pt]{cip-submit-2019}  
\usepackage{color,soul}
\usepackage{afterpage}

\usepackage{wrapfig}
\usepackage{wasysym}
\usepackage{enumitem}
\usepackage{adjustbox}
\usepackage{ragged2e}
\usepackage[svgnames,table]{xcolor}
\usepackage{tikz}
\usepackage{longtable}
\usepackage{changepage}
\usepackage{setspace}
\usepackage{hhline}
\usepackage{multicol}
\usepackage{tabto}
\usepackage{float}
\usepackage{multirow}
\usepackage{makecell}
\usepackage{mathptmx}
\DeclareSymbolFont{epsilon}{OML}{cmm}{m}{it}
\DeclareMathSymbol{\epsilon}{\mathord}{epsilon}{"0F}
\def\DD{D\kern-.7em\raise0.25ex\hbox{\char '55}\kern.33em}
\usepackage{hyperref}
\hypersetup{
	colorlinks=false,
	pdfborder={0 0 0}
}
\usepackage{graphicx,subfigure,wrapfig,epstopdf}
\def\Mr{\uppercase}
\usepackage{graphics}
\usepackage{wasysym}
\usepackage{amsmath,amsxtra,amssymb,latexsym,amscd,color,cite,bm}
\usepackage[mathscr]{eucal}
\usepackage{array}
\usepackage{multirow}
\usepackage{cite} 

\def\doi#1{\href{http://dx.doi.org/#1}{doi:#1}}
\def\titles#1{\title{\large\bf\noindent #1}}
\def\authors#1{\author{\begin{flushleft}{#1}\end{flushleft}}}
\def\authord#1#2{\indent\Mr{#1}$^{#2}$}
\def\addressed#1#2{\\[1mm]\textit{$\!\!\!^{#1}$\indent#2}}

\def\Email{$^{\dagger}$}
\def\email#1{\bigskip\href{mailto:#1}{\!\!\Email\textit{E-mail:}~{#1}}\\[3mm]}
\def\PublicationInformation#1#2#3#4{\\[4mm]\href{mailto:#1}{\!\!\Email\textit{E-mail:}~{#1}}\\[3mm]
	\textit{\indent Received #2}\\[1mm]
	\textit{Accepted for publication~#3}\\[1mm]
	\textit{Published~#4}}
\def\Keywords#1{\\[.2cm] \textnormal{Keywords:~{#1}}.} 
\def\AND{$\text{\Small AND }$}
\def\and{$\text{\tiny AND }$}

\def\Classification#1{\\[.2cm] \textnormal{Classification numbers:~{#1}.}}  

\newcommand{\Msun}{M$_{\odot}$}
\newcommand{\kms}{km\,s$^{-1}$\,}
\newcommand{\kmsns}{km\,s$^{-1}$}
\newcommand{\dego}{$^{\circ}\,$}
\newcommand{\degons}{$^{\circ}$}

\newcommand{\sout}[1]{\unskip}

\def\ed{
	\bibliographystyle{cip-sty-2019}
	\bibliography{references-database-name}
\end{document}
}

\begin{document}
	\Year{2020}
	\Page{1}\Endpage{20}
	\titles{Morpho-kinematics of the molecular gas in a quasar host galaxy at redshift $z$=0.654}
	\authors{
	\authord{T.T.Thai}{1,2}, \authord{P. Tuan-Anh}{1}\Email, \authord{P. Darriulat}{1}, \authord{D. T. Hoai}{1}, \authord{P. T. Nhung}{1}, \authord{P. N. Diep}{1}, \authord{N. B. Ngoc}{1}, \AND \authord{N. T. Phuong}{1}
	\newline
	\addressed{1}{Vietnam National Space Center, Vietnam Academy of Science and Technology\\
	18 Hoang Quoc Viet, Cau Giay, Hanoi, Vietnam}
	\addressed{2}{Graduate University of Science and Technology\\
			18 Hoang Quoc Viet, Cau Giay, Hanoi, Vietnam}
\PublicationInformation{ptanh@vnsc.org.vn}{\today}{??}{??}
	}
	\maketitle
	\markboth{T.T.Thai, P. Tuan-Anh \& P. Darriulat}{Morpho-kinematics of quasar host galaxy RXJ 1131}

\begin{abstract}
We present a new study of archival ALMA observations of the CO(2-1) line emission of the host galaxy of quasar RX J1131 at redshift $z$=0.654, lensed by a foreground galaxy. A simple lens model is shown to well reproduce the optical images obtained by the Hubble Space Telescope. Clear evidence for rotation of the gas contained in the galaxy is obtained and a simple rotating disc model is shown to give an excellent overall description of the morpho-kinematics of the source. The possible presence of a companion galaxy suggested by some previous authors is not confirmed. Detailed comparison between model and observations gives evidence for a more complex dynamics than implied by the model. Doppler velocity dispersion within the beam size in the image plane is found to account for the observed line width.
\Classification{
  {\textcolor{blue}{98.65-r}}}
\end{abstract} 

\section{\Mr{Introduction}}\label{sec1}
\subsection{General features}

RX J1131-1231 (simply called RX J1131 in the following), is a distant quasar, at redshift $z_s\sim$0.654, corresponding to a distance of $\sim$1.45 Gpc or a time of $\sim$7.5 Gyr after the Big Bang, about half way from us, and some 4 Gyr later than the time of maximal star formation \cite{Adam+etal+2016, Ade+etal+2016}. At such distance, 1 arcsec spans 7.03 kpc. It hosts a Super Massive Black Hole (SMBH) in its centre with a mass of $\sim$2 10$^8$ \Msun; it rotates extremely fast, reaching near half the light velocity \cite{Reis+etal+2014}. The quasar and its host galaxy are gravitationally lensed by a galaxy in the foreground, at redshift $z_L\sim$0.295. They are the object of numerous studies, in particular aiming at a better understanding of the cosmological parameters governing the expansion of the Universe (\cite{Tewes+etal+2013, Bonvin+etal+2017} and references therein). Microlensing caused by stars transiting across the line of sight to the quasar has been used to study the structure of the lens halo (\cite{Chartas+etal+2012, Sugai+etal+2007} and references therein).\\
\begin{figure*}[!h]
  \includegraphics[width=1.\textwidth,trim=0cm 0cm 0.cm 0.cm,clip]{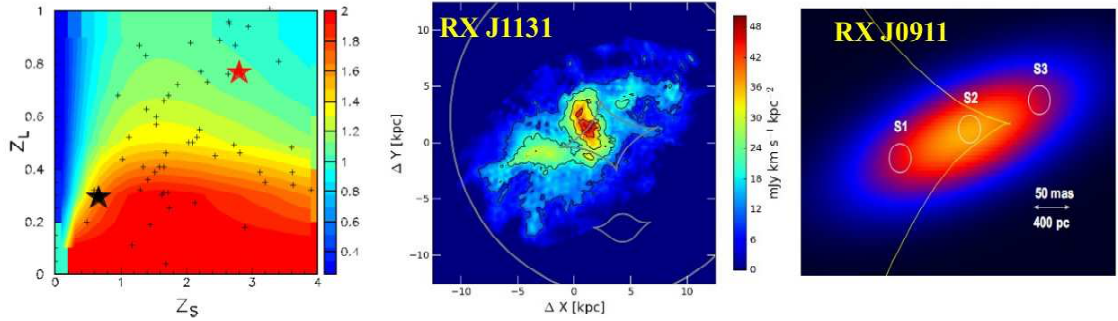}
  \caption{Left: dependence on the redshifts of the source ($z_S$ in abscissa) and of the lens ($z_L$ in ordinate) of the ratio between their respective angular diameter distances $d_{aS}$/$d_{aL}$ \cite{Adam+etal+2016, Ade+etal+2016}). The relative size of the lens with respect to the source is proportional to $d_{aS}$/$d_{aL}$. The stars show the locations of quasars RX J1131 (black, P18) and RX J0911 (red, ~\cite{Anh+etal+2013}, ~\cite{Anh+2014}). The sizes of the host galaxies are compared to the size of the lens caustic in the central and right panels respectively.}
  \label{fig1}
\end{figure*}
Infrared observations obtained by Herschel \cite{Stacey+etal+2018} have measured the spectral energy distribution (SED), and archival VLA observations (Program ID: AW741; PI: Wucknitz) analysed by \mbox{Leung et al. 2017} \mbox{\cite{Leung+etal+2017}}, referred to as L17 in the following, have shown resolved continuum emission from the jets and the core of the foreground elliptical galaxy as well as emission toward the background quasar.

Thorough analyses of high angular resolution HST optical and NIR images \cite{castles,Claeskens+etal+2006, Birrer+etal+2016} and of Keck Adaptive Optics images  \cite{Chen+etal+2016} have produced a detailed description of the lensing properties in the neighbourhood of the quasar. They reveal a typical long axis quad configuration \cite{Blandford+etal+1989, Saha+2003}, the quasar being located within the eastern cusp of the lens caustic. As emission from the lens galaxy is simultaneously detected, the parameters of the lensing potential can be accurately evaluated. However, they probe only the vicinity of the cusp of the caustic curve. As the emission of the quasar host galaxy covers the whole caustic curve and extends even farther out, one cannot take it as granted that the simple lens model obtained from the study of the quasar images reliably applies to the whole host galaxy. This is at variance with the gravitational lensing of quasar hosts that are farther away and cover only part of the caustic in addition to being intrinsically smaller. The central region of the caustic corresponds to images close to an Einstein ring configuration, which dominates the picture in the case of RX J1131. We illustrate this feature in \mbox{Figure \ref{fig1}}, which shows the location of RX J1131 in the plane $z_L$ vs $z_S$, lens vs source redshifts, together with that of other multiple imaged systems \cite{castles}, and compares it with the case of a typical farther away quasar host galaxy, RX J0911 ~\cite{Anh+etal+2013, Anh+2014}.

\subsection{Millimeter observations}
The present work uses archival ALMA observations to study the emission of the \mbox{CO(2-1)} molecular line of the quasar host galaxy, with a beam of $\sim$0.4$\times$0.3 arcsec$^2$ providing unprecedented image quality. The data have been analysed in much detail by the team who proposed the observation (\mbox{\cite{Paraficz+etal+2018}}, referred to as P18 in the following) and who kindly sent us data files summarizing the results of their analysis.
The CO(2-1) line emission of RX J1131 was observed by L17 at the Plateau de Bure Interferometer (PdBI) with an angular beam size (FWHM) of 4.4$\times$2.0 arcsec$^2$, a spectral resolution of $\sim$21.5 \kms and a noise level of $\sim$1.45 mJy beam$^{-1}$ per channel. In addition to their CO(2-1) PdBI observations, L17 used the Combined Array for Research in Millimeter-wave Astronomy (CARMA) to detect the \mbox{CO(3-2)} line emission but the signal to noise ratio is at only 5-$\sigma$ significance.

ALMA observations of both 2 mm continuum and CO(2-1) line emission have been analysed by P18 with an angular resolution of $\sim$0.3 arcsec, an order of magnitude better than for the PdBI data of L17. These are the data used in the present article, about which P18 kindly sent us useful documentation that complements the published article (private communications by Professors Frederic Courbin and Matus Rybak).

\begin{figure*}[!h]
  \includegraphics[width=1.\textwidth,trim=0cm 0cm 0.cm 0.cm,clip]{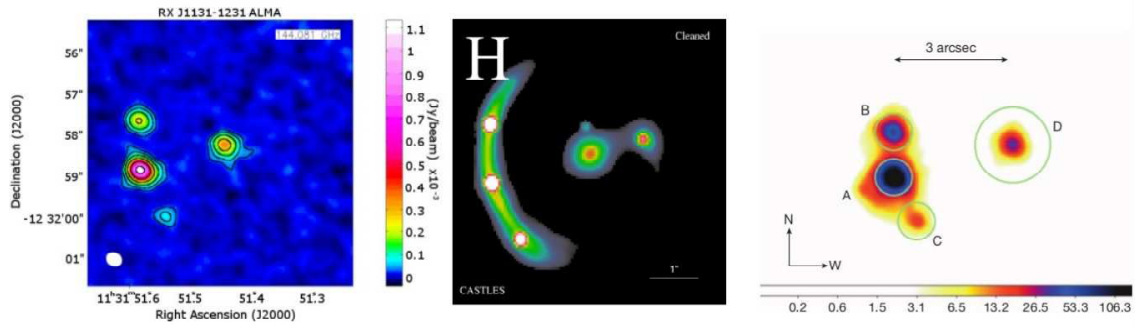}
  \caption{Images of RX J1131. Left:  ALMA 2 mm continuum (P18). Centre: HST visible (CASTLES). Right: CHANDRA, 0.3 to 8 keV X-rays \cite{Reis+etal+2014}.}
  \label{fig2}
\end{figure*}

The continuum image obtained by P18 shows four clearly separated compact components, three coincident with the lensed optical point images and one associated with the lens galaxy (Figure ~\ref{fig2} left). In contrast with continuum emission, CO(2-1) line emission detects no signal from the lens galaxy. The velocity integrated intensity map clearly shows (76-$\sigma$) line emission extended over a complete Einstein ring. The map of velocity dispersion displays values covering from $\sim$10 to $\sim$50 \kmsns. The authors of P18 note that peaks in the intensity map and the region of high velocity dispersion are coincident and probably reveal on-going star-formation; these peaks are not strictly coincident with the quasar emission.

As mentioned above, one cannot take it as granted that the simple lens model obtained from the study of the quasar images reliably applies to the whole host galaxy. For this reason P18 reconstruct the source brightness distribution using the ALMA CO(2-1) data exclusively \cite{Rybak+etal+2015a, Rybak+etal+2015b, Vegetti+etal+2009}. The result (Figure \ref{fig1}, central panel) is consistent with a large rotating disc inclined by 54\dego with respect to the plane of the sky, having rotation velocity reaching over 400 \kmsns.

The model used by P18 to describe the lens includes an external shear, ellipticity of the main lens and a contribution from the small satellite galaxy revealed by the HST image north of the main lens (central panel of Figure \ref{fig2}). The contribution of the latter is known to be very small \cite{Claeskens+etal+2006} and the contribution of ellipticity, while different from that of an external shear, does not affect strongly the general picture: excellent descriptions of the lensing of the quasar point source are obtained with either external shear alone or ellipticity alone and their combination in the P18 model causes a rotation of only 4\dego of the caustic with respect to a model ignoring ellipticity. For this reason our approach in the present article is to use a simple lens potential with only external shear and no ellipticity to describe the lensing of both the quasar point source and its host galaxy. This has the advantage of producing a lens equation that can be solved analytically and lends itself to simple and transparent interpretations.

The aim of the present article is to shed a new light on the results obtained by P18 by evaluating uncertainties attached to their main results. To this end we use a different method of data reduction, resulting in a better angular resolution but an accordingly higher noise level; we work with the clean image in the plane of the sky rather than in the uv plane as P18 do; we use a simpler description of the lensing mechanism based exclusively on the analysis of the quasar point source images. Altogether, this simpler approach has the advantage of transparency and of lending itself easily to interpretation. It has no pretention for being better than the approach used by P18; on the contrary, working in the uv plane allows for a more reliable treatment of noise than working on the clean sky plane image. But we show that it gives as proper a description of the main results, which justifies its use. Details of the analysis can be found in \mbox{Thai T. T. 2020\cite{Thai+2020}}; here, the emphasis is on presenting and discussing the main results.   
Sections \ref{sec2} and \ref{sec3} describe the lens model and present its results both for the quasar point source (Section \ref{sec2}) and for the molecular gas emission (Section \ref{sec3}). They are compared with the results obtained by other authors and some issues, such as the possible presence of a companion of the quasar host galaxy, are addressed in this context. 
Section \ref{sec4} compares the data with the prediction of a simple rotating thin disc model and offers a detailed discussion of the relative contributions of turbulence and rotation to the observed Doppler velocity spectra. A summary and conclusions are presented in Section \ref{sec5}.

\section{The quasar point source: a simple lens model} \label{sec2}
\subsection{Lens equation}\label{sec2.1}
The knowledge of the position and luminosity of the point images of the quasar, together with the direct observation of point-like emission from the lens, allow for an accurate evaluation of the effective lensing potential that describes the lens in the vicinity of the quasar. The HST image positions measured with respect to the lens G are measured with a mean precision of $\sim$8 mas. The deflection induced by the lens as a function of the sky coordinates of the image can be described by an effective potential $\psi$ proportional to the integral of the gravity potential along the line of sight between source and observer. Convenient forms include an elliptical lens and/or an external shear \cite{Blandford+etal+1989, Saha+2003,  Hoai+etal+2013}. Claeskens et al. 2006 \cite{Claeskens+etal+2006}, using such potentials, found that both give excellent results, the position angles of the minor axis of the ellipse and of the external shear being identical, $\sim$16\dego east of north. This shows that introducing a shear or an ellipticity is an ad hoc way to model the anisotropy of the lens mass distribution, mostly due to the presence of a massive cluster of galaxies distant by a few arcminutes in the north-eastern direction \cite{Suyu+etal+2013}. Accordingly, \mbox{we choose to use an effective potential of the form}
\begin{equation}\label{rela1}
  \psi=r_0r+\frac{1}{2}\gamma_0 r^2\cos2(\varphi-\varphi_0).	      
\end{equation}
where $r$ and $\varphi$ (measured counter-clockwise from west) are polar coordinates of the point image with origin at the centre of the lens galaxy.  
The first term describes a spherical main lens of Einstein radius $r_0$. The second term represents a shear of strength $\gamma_0$ at position angle $\varphi_0$. Writing that the gradient of the potential cancels, and calling ($r_s$, $\varphi_s$) the polar coordinates of the point source, one obtains the lens equation: $r_se^{i\varphi_s}=re^{i\varphi}(1-r^{-1}\frac{\partial\psi}{\partial r}-ir^{-2}\frac{\partial\psi}{\partial\varphi})$.
For the potential of Relation (\ref{rela1}), writing separately the real and imaginary parts, one obtains:
\begin{align}\label{rela2}
&r_s\cos(\varphi_s-\varphi)= r(1-\gamma_0\cos2(\varphi-\varphi_0))-r_0=A \nonumber \\
&r_s\sin(\varphi_s-\varphi)= r\gamma_0\sin2(\varphi-\varphi_0)=B.        		   \end{align}  
Relations (\ref{rela2}) can be used to simply obtain the position of a point source from that of a point image:
\begin{align}\label{rela3}
  r_s=(A^2+B^2)^{\frac{1}{2}}\, \, \, \text{and} \, \, \, \varphi_s=\varphi+\tan^{-1}(B/A). 
\end{align}
	The first of these relations can be rewritten as \mbox{$r_s^2=(r-r_0)^2+r^2\gamma_0^2-2r\gamma_0(r-r_0)\cos2(\varphi-\varphi_0)$} implying that $r_s$, $r\gamma_0$ and $|r-r_0|$ form a triangle with an angle $2(\varphi-\varphi_0)$ facing $r_s$.  
Imaging a point source is done by eliminating $r$ from Relations (\ref{rela2}). One then obtains an equation in $\varphi$ giving four images when the source is inside the caustic and two when it is outside. \mbox{The image magnification is obtained by differentiating the lens equation}.
\begin{figure*}[!h]
  \includegraphics[width=1.\textwidth,trim=0cm 0cm 0.cm 0.cm,clip]{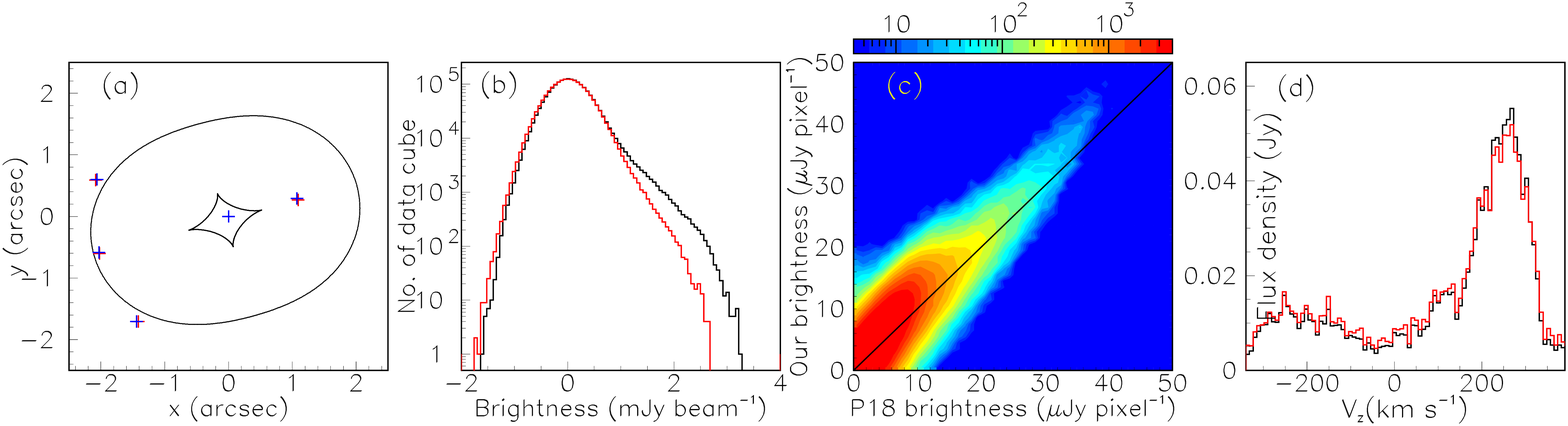}
  \caption{a) Best fit results comparing observations (blue) and model (red). The semi-major and minor axes of the caustic are 589 and 446 mas respectively; those of the critical curve are 2.14 and 1.62 arcsec. b) Observed brightness distribution (mJy beam$^{-1}$) in the image plane (black for P18 and red for our data). c) Correlation between P18 (abscissa) and our (ordinate) brightness measurements. d) Doppler velocity spectra for P18 (black) and our (red) data after application of a 0.45 mJy cut on large pixels (\mbox{250$\times$250 mas$^2$)}.}
  \label{fig3}
\end{figure*}

\subsection{Results}\label{sec2.2}
Seven parameters are adjusted by optimizing the match between the observed quadruple HST point images and the prediction of the model: three parameters ($r_0,\gamma_0,\varphi_0$) define the lensing potential; two account for a possible offset of the lens centre ($\Delta x$, $\Delta y$) with respect to the emission of the lens galaxy (G); and two (\mbox{$r_s$, $\varphi_s$}) locate the point source with respect to the lens centre: its coordinates with respect to G are therefore $x_s$+$\Delta x$ and $y_s$+$\Delta y$ with $x_s=r_s\cos\varphi_s$ and $y_s$=$r_s\sin\varphi_s$. We minimize the value of the root mean square deviation $\delta$ between model and observation. The best fit gives $\delta$=14 mas and is illustrated in Figure ~\ref{fig3}a. The best fit values of the model parameters are $r_0$=1.84$\pm$0.02 arcsec, $\gamma_0$=0.138$\pm$0.007, $\varphi_0$=106\dego$\pm$1\dego, $x_s$=$-$0.47$\pm$0.03 arcsec, $y_s$=$-$0.14$\pm$0.01 arcsec, $\Delta x$=$-$50$\pm$167 mas and $\Delta y$=$-$60$\pm$20 mas in excellent agreement with the results obtained by Claeskens et al. 2006 \cite{Claeskens+etal+2006}. The quoted uncertainties are arbitrarily defined as doubling the value of $\delta^2$. We find strong correlations between the model parameters due to the fact that what is measured accurately is the relative position of the source with respect to the cusp of the caustic, not with respect to its centre. Magnifications cannot be reliably calculated because of the effect of microlensing ~\cite{Sluse+etal+2006, Chen+etal+2016} and are not used here.

In summary a lens model using the effective potential of the form given by Relation (\ref{rela1}) gives an excellent description of the astrometry of the HST images. Agreement has been obtained with the results of earlier analyses and a good understanding of the uncertainties attached to the model parameters and of their correlations has been reached. However, these results probe only the environment of the eastern cusp of the caustic and the validity of the model at larger distances, in the region covered by the emission of the host galaxy, cannot be taken as granted.

\section{The host galaxy: de-lensing the observed emission of the CO(2-1) line} \label{sec3}
\subsection{Data reduction}
We use ALMA observations, project number 2013.1.01207.S (PI: Paraficz Danuta), collected on July 19th 2015 using the normal 12-m array with 37 antennas covering baselines between 27.5 m and 1.6 km. Details of the observations and of the data reduction are given in \mbox{Thai T. T. 2020 \cite{Thai+2020}}. Imaging was performed using the standard CLEAN algorithm applied to the calibrated visibilities. With the aim to understand the effect of a different data reduction than used by P18 on the results obtained, we used robust weighting rather than natural weighting as adopted by P 18. This means a better angular resolution (the beam is 70\% in area compared with the P18 beam) but a larger noise level. This dictated our choice of 0.5 as robust parameter, a reasonable compromise between angular resolution and noise. We recall that a rigorous treatment of the noise is only possible in the uv plane. Here, as we work in the image plane, we can only obtain an approximate estimate of its level. However, we have been careful to take this caveat in due account whenever relevant to the argument being made. Continuum emission is found to agree precisely with the results obtained by P18 and is not discussed in the present article.
We imaged the CO(2-1) data in the form of a cube of 640$\times$640 pixels, each 70$\times$70 mas$^2$, covering a square of $\pm$22.4 arcsec centred on the continuum emission of the lens galaxy $-$ G $-$ and of 121 Doppler velocity bins, \mbox{8.417 \kms each}, covering an interval of $\pm$509 km s$^{-1}$, centred on the red-shifted ($z$=0.654) frequency of the CO(2-1) line emission. The beam size is \mbox{380$\times$290 mas$^2$} with position angle of 66\dego east of north; the noise rms level is 0.38 mJy beam$^{-1}$ per channel. In the remaining of the article, we use coordinates centred at the best-fit lens centre, 60 mas south and 50 mas east of G, with the $y$ axis pointing 16\dego east of north and the $x$ axis pointing 16\dego north of west, perpendicularly to the external shear (namely $\varphi_0$=90\dego in Relations \ref{rela1} to \ref{rela3}). In this new frame, using the axes of the caustic and critical curve as axes of coordinates, the quasar is located at $x_s$=$-$0.49 arcsec and $y_s$=$-$0.005 arcsec. 
\subsection{Observed emission of the CO(2-1) line}
	Comparing the CO(2-1) data reduced above with the P18 data reveals the differences in beam size (\mbox{380$\times$290 mas$^2$} instead of 440$\times$360 mas$^2$) and noise level (3.0 instead of 1.9 mJy arcsec$^{-2}$). The comparison is made in a square of 6.25 arcsec side centred on the lens centre, containing 125$\times$125 pixels, each \mbox{50$\times$50 mas$^2$}. Eight different data sets are considered separately, each covering an 84.17 \kms interval of Doppler velocity.  While the brightness distributions of the two sets are similar when expressed in \mbox{mJy beam$^{-1}$} (Figure \ref{fig3}b), they are scaled relative to each other by the ratio of the beam area when expressed in mJy pixel$^{-1}$, mJy arcsec$^{-2}$. The correlation between the two sets of brightness measurements is illustrated in Figure ~\ref{fig3}c. Applying a common cut of 10 $\mu$Jy per pixel (4 mJy arcsec$^{-2}$), which suppresses much of the noise, gives a good agreement between the brightness measurements of the two data sets: on average, the asymmetries (difference divided by sum) between the brightness integrated over each of the 8 velocity intervals cancel and have a root mean square deviation of 5\%. The effect of the different noise levels is attenuated when using larger pixels: Figure ~\ref{fig3}d compares the Doppler velocity spectra obtained by applying a cut of 0.45 mJy per pixel of 250$\times$250 mas$^2$ (7.2 mJy arcsec$^{-2}$). The differences between P18 and our brightness measurements give a measure of the uncertainties resulting \mbox{from differences in data reduction.}
\subsection{De-lensing}
\begin{figure*}[!h]
  \centering
  \includegraphics[width=.7\textwidth,trim=0cm 0.5cm 0.2cm 0.cm,clip]{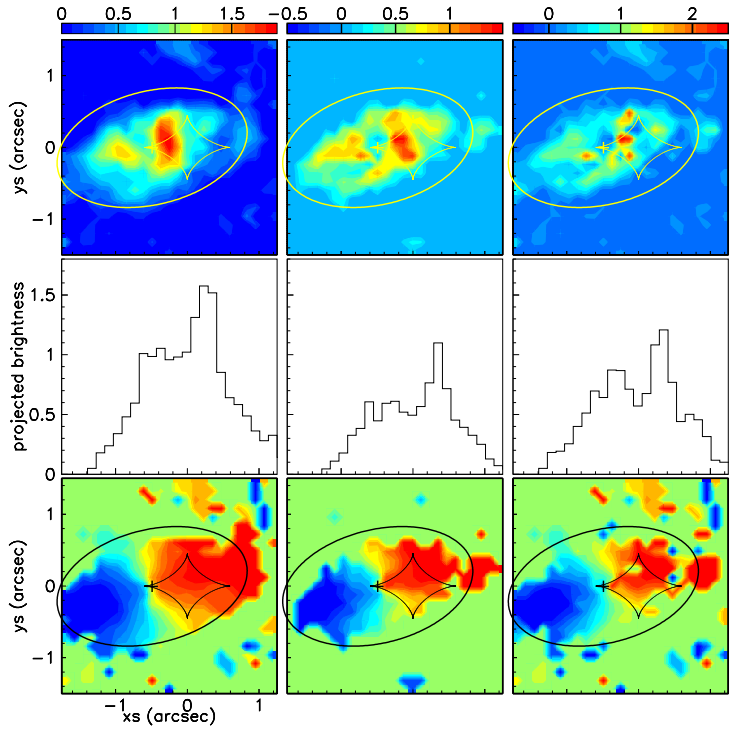}
  \caption{Upper row: maps of the intensity, integrated over the whole velocity range, of the source emission as obtained by P18 (left), by direct de-lensing (centre) and by $\chi^2$ minimization (right). The location of the quasar is indicated by a cross. Units are Jy \kms arcsec$^{-2}$.  Ellipses show the projection of the model disc. Middle row: projections of the source brightness (Jy \kms arcsec$^{-1}$) on the major axis of its elliptical projection on the sky plane ($x^\prime$ axis in Figure ~\ref{fig6}); only pixels located inside this ellipse are included. Lower row: maps of the mean Doppler velocity.}
  \label{fig4}
\end{figure*}

We reconstruct the CO(2-1) emission in the source plane using the simple lens model defined in \mbox{Section \ref{sec2.2}} with $r_0$=1.84 arcsec and $\gamma_0$=0.138. We consider separately 8 Doppler velocity intervals, each \mbox{84.17 \kms} wide, covering between $-$340 \kms and +333 \kmsns. We use two different methods to de-lens the observed images. One is the simple direct de-lensing described by Relations (\ref{rela3}) in Section \ref{sec2.1}, which has the advantage of simplicity and of inviting transparent interpretations. The method has two drawbacks. It lacks control of the effect of beam convolution, the de-lensed source brightness being smeared by the beam in a way that depends on the location of the image pixel. And it introduces de-lensing noise, implying the application of a strong cut on image brightness in order to stand aside from it. We have been careful to ensure that our results were robust with respect to both. For this reason one usually prefers to start from a model of the source brightness, image it, convolve it with the beam and compare the result with the observed image (as we do with the second method). In practice, application of the first method requires probing each image pixel over its whole area of 50$\times$50 mas$^2$: we use 1000 random points per pixel containing brightness in excess of \mbox{6 $\mu$Jy pixel$^{-1}$} and take a proper weighted average in each source pixel of the de-lensed values obtained for the brightness. Taking such a proper weighted average is not trivial. We need to know, for each image pixel, if it is imaged by a lens producing 2 images (outside the caustic) or 4 images (inside the caustic). In the first case a weighted average of the de-lensed brightness gives the source brightness outside the caustic and in the second case another weighted average of the de-lensed brightness gives the source brightness inside the caustic. These two weighted averages need to be evaluated separately. Results are displayed in Figure ~\ref{fig4}. The second method proceeds in the opposite direction, from source to image. Imaging is done using a matrix of elements $f_{ijkl}$ equal to the brightness obtained in image pixel ($k,l$) by lensing source pixel ($i,j$) of unit brightness ($f_{ijkl}$ is a pure number). The matrix has been calculated once for all and includes the effect of beam convolution. We use 25$\times$25 source pixels of 120$\times$120 mas$^2$ each, making a square covering \mbox{3$\times$3 arcsec$^2$}. Optimization is made by minimizing the value of $<$$\delta O^2$$>$, the mean square deviation between observed and modelled brightness in the image pixels. We use large pixels in the image plane, \mbox{250$\times$250 mas$^2$}. We simply loop over the source pixels containing brightness in excess of a threshold of $\sim$0.3 mJy arcsec$^{-2}$, vary their brightness by a small quantity $\pm \epsilon$ and calculate the new value of $<$$\delta O^2$$>$. The iteration uses the P18 source brightness as a first approximation. We retain as new value of the brightness the value giving the smaller value of $<$$\delta O^2$$>$. We repeat the procedure until all pixels contain a brightness $S$ giving a smaller value of $<$$\delta O^2$$>$ than for $S\pm \epsilon$. Convergence is achieved after 60 to 140 iterations depending on the velocity interval.

\begin{figure*}[!h]
  \centering
  \includegraphics[height=3.6cm,trim=0.cm 0.5cm 0.05cm 0.cm,clip]{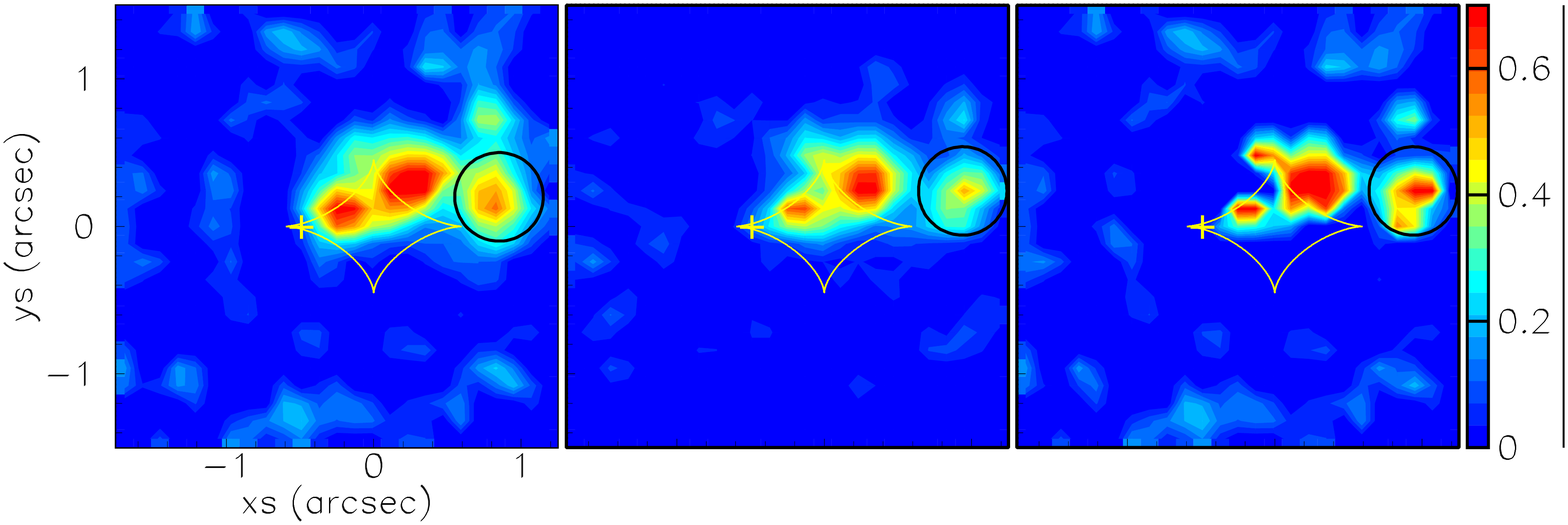}
  \includegraphics[height=3.6cm,trim=0.cm 0.5cm 0.05cm 0.cm,clip]{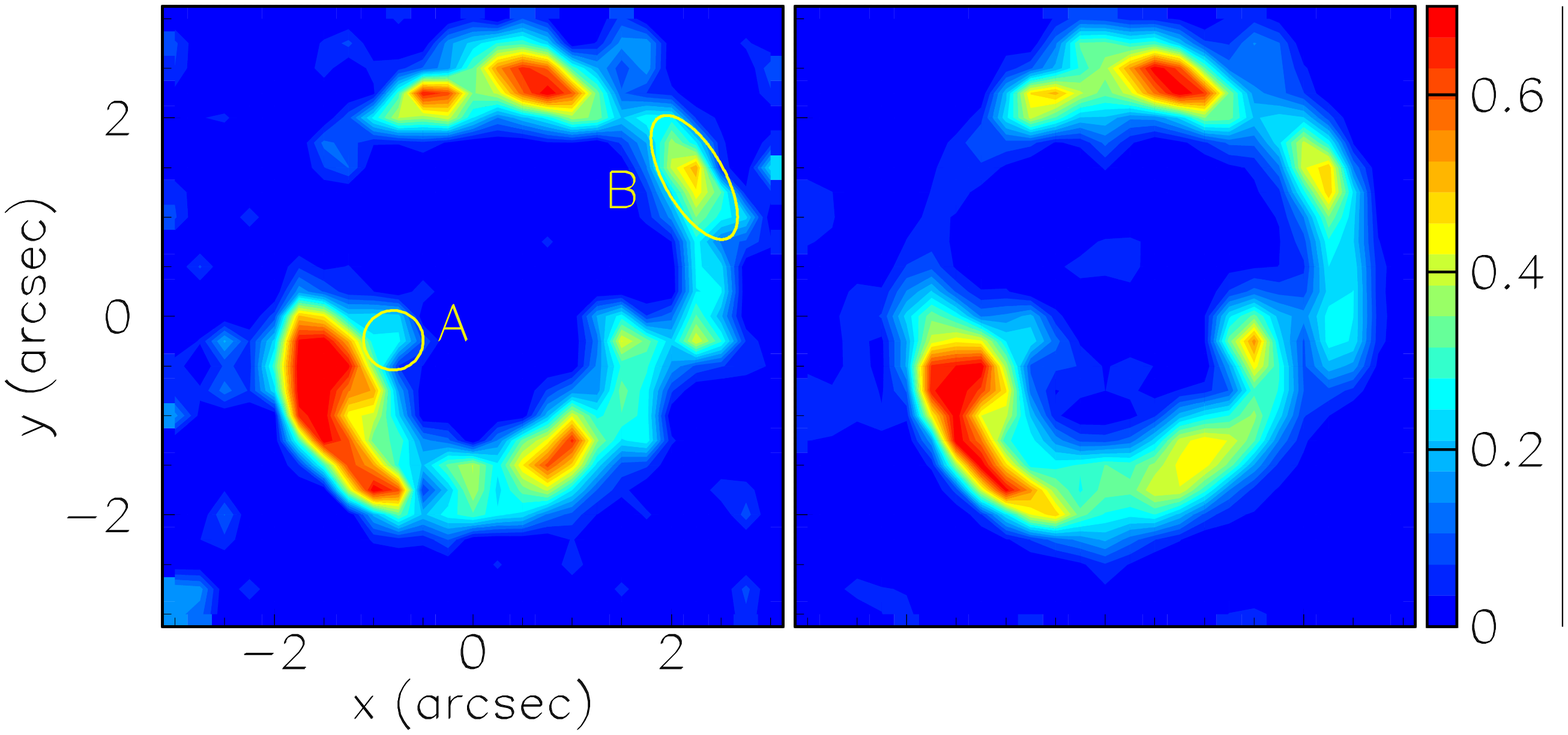}
\caption{Maps of the brightness integrated over the red-most velocity interval (249 to 333 \kmsns) are shown in the three leftmost panels for the source and in the two rightmost panels for the image. The source maps are, from left to right, for P18, for direct de-lensing and for our best-fit result. The image maps show, again from left to right, the observed images and the best-fit results. Colour scales are in Jy \kms arcsec$^{-2}$.}
 \label{fig5}
\end{figure*}

\mbox{Figure ~\ref{fig4}} maps the source brightness integrated over the whole velocity range and the mean Doppler velocity, together with those obtained by P18 and by direct de-lensing. The three intensity maps of the source emission are consistent with the elliptical projection on the sky plane of a thin circular disc centred on the quasar ($x$=$-$0.49 arcsec, $y$=$-$0.005 arcsec). We find that the major axis of the ellipse of projected emission is oriented some 14\dego north of the $x$ axis, meaning 30\dego north of west (L17 quote a value of 31\degons). We evaluate the lengths of the major and minor axes to be $\sim$2.7 and $\sim$1.6 arcsec ($\sim$19 and $\sim$11 kpc respectively) corresponding to an inclination of the disc with respect to the plane of the sky of cos$^{-1}$(1.6/2.7)=54\dego as obtained by P18. We note however that the long axis of the P18 caustic is only $\sim$12\dego south of east instead of 16\dego in our simple lens model. Also shown in Figure ~\ref{fig4} are projections of the intensity on the major axis of the ellipse. All three distributions, rather than peaking at the quasar location, display a small depletion in its vicinity. The maps of the mean Doppler velocity display a strong gradient along the major axis, as expected from a rotating thin disc. L17 have claimed evidence for the presence of a companion of the quasar host galaxy in the red-most velocity interval covering from 249 to 333 \kmsns. This companion is supposed to match approximately image F of \mbox{Brewer \& Lewis 2008 \cite{Brewer+2008}}, some 100 to 200 mas south-west of the western cusp of the caustic. We find indeed an enhancement of emission in the red-most velocity interval, centred at $x\sim$0.8 arcsec and \mbox{$y\sim$0.2 arcsec} (Figure ~\ref{fig5}). It is most probably what L17 are referring to, but it does not match precisely image F of \mbox{Brewer \& Lewis 2008 \cite{Brewer+2008}}, being north rather than south of the western cusp of the caustic. Lensing this lump produces two images, shown as A and B in Figure ~\ref{fig5}. Contrary to image B, image A stands out of the general image morphology; it has \mbox{a magnification of $\sim$0.5} \mbox{while image B has a magnification of $\sim$4.3}. While much weaker than image B, image A contributes as much as image B to the de-lensed source brightness and is therefore responsible for the appearance as an isolated lump in the source plane: the case for a lump of separate emission is much weaker when made in the image plane than in the source plane. The fact that the Doppler velocity of the lump matches well that implied in this region by the galaxy rotation curve argues against the L17 interpretation as a companion galaxy. It is more natural to interpret it as a lump of enhanced emission on the disc.
\section{A simple rotating disc model} \label{sec4}
\subsection{Geometry}
\begin{figure*}[!h]
  \centering
  \includegraphics[height=5cm,trim=-1cm -2.5cm 0.cm -1.1cm,clip]{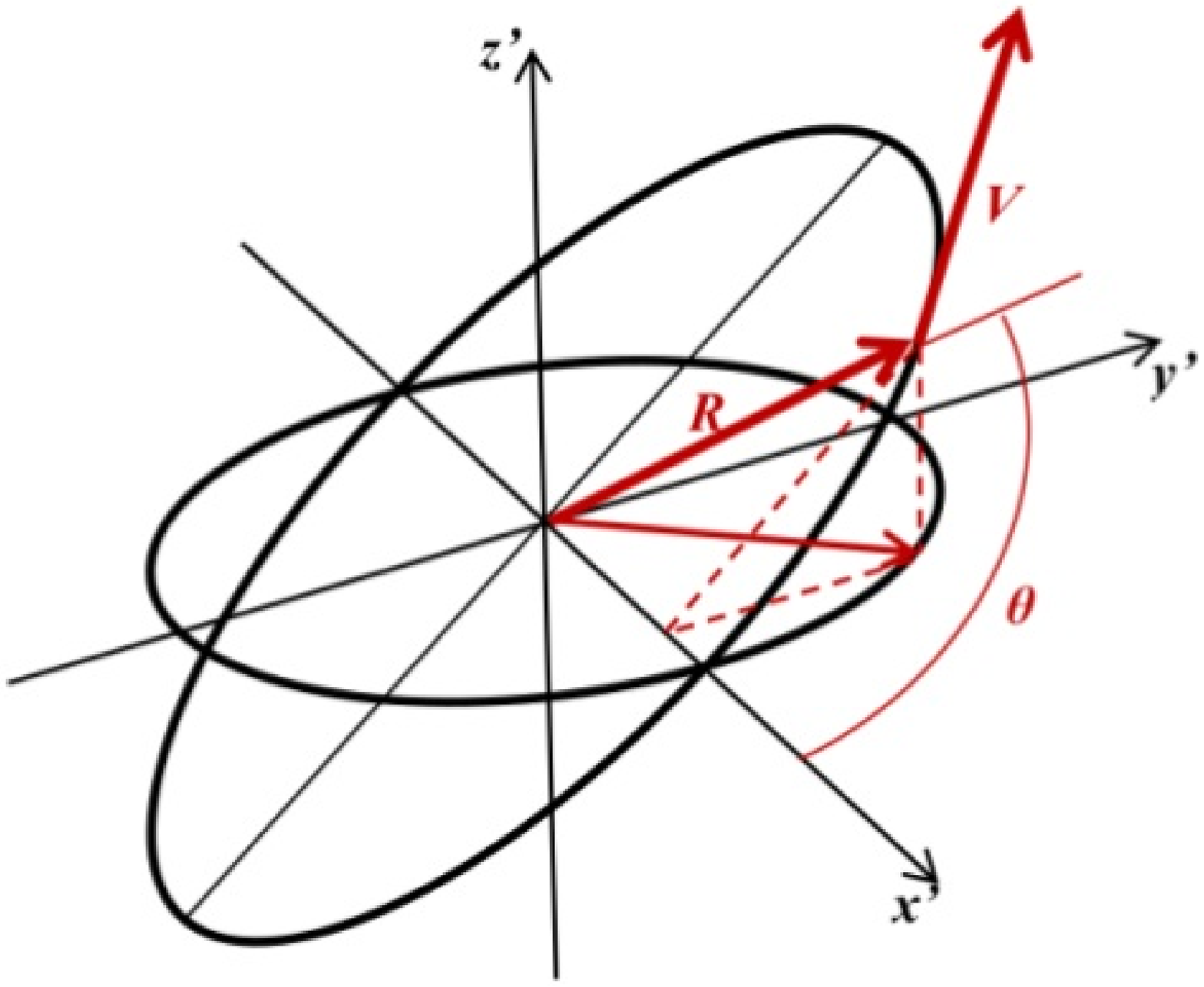}
  \includegraphics[height=5cm,trim=0cm 0.5cm .05cm 0.1cm,clip]{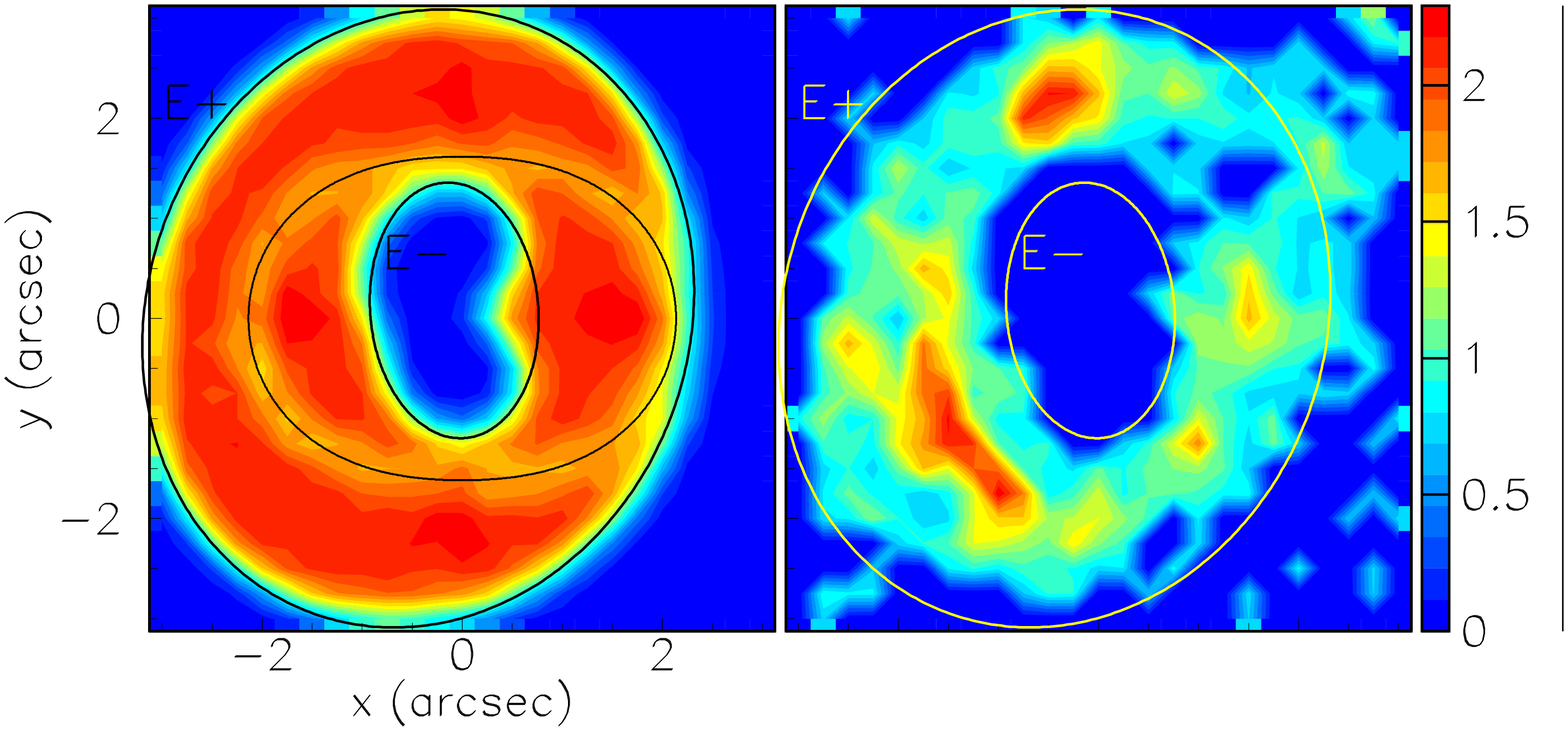}
\caption{Left: geometry in a system of coordinates ($x^\prime$,$y^\prime$,$z^\prime$) centred on the quasar with $x^\prime$ along the trace of the disc on the plane of the sky and $z^\prime$ perpendicular to the plane of the sky. Centre and right: the image of a disc of uniform brightness lensed by the simple lens potential (centre) is compared with observation (right). Units are arbitrary for the model and Jy \kms arcsec$^{-2}$ with a cut at \mbox{0.67 Jy \kms arcsec$^{-2}$ for the observed data}.}
 \label{fig6}
\end{figure*}

The preceding sections have shown that the general morpho-kinematics of the de-lensed images is a robust result of the analysis of P18. Using a simpler lens does not strongly affect the brightness distribution in the source plane and preserves the Doppler velocity distribution, typical of a thin rotating disc inclined with respect to the plane of the sky. In order to have a reference with which one can make quantitative comparisons, it is instructive to construct the image produced by a rotating disc of uniform brightness having morpho-kinematics matching the distributions displayed in Figure ~\ref{fig4}. The geometry is illustrated in the left panel of Figure ~\ref{fig6}.

Defining the position of a point in the disc by its polar coordinates ($R$, $\theta$) with $\theta$=0 along $x^\prime$, \mbox{intersection of the disc with the plane of the sky containing the quasar, we see from Figure ~\ref{fig6} that}
\begin{align}\label{rela4}
  &x^\prime=R\cos\theta \,\,\,\,\,\,\,\,\,\,\,\, y^\prime=R\sin\theta \cos54^\circ \,\,\,\,\,\,\,\,\,\,\,\,\,\, z^\prime=R\sin\theta\sin 54^\circ \nonumber\\
  &V_x=-V(R)\sin\theta \,\,\,\,\,\,\,\,\,\,\, V_y=V(R)\cos\theta\cos54^\circ  \,\,\,\,\,\,\,\,\,\, 
  V_z=V(R)\cos\theta\sin54^\circ .
\end{align}
In our system of coordinates, with $x$ pointing 16\dego north of west, the $x^\prime$ axis points to a direction of 30\dego$-$16\dego=14\dego; 
centring the disc on the quasar of coordinates ($x,y$)=($-$0.49, $-$0.005) arcsec, we obtain:
\begin{align*}
 &\,\,\,\,\,\,\,\,\,\,\,\,\,\,\,\,\,\,\,\,\,\,\,\,\,\,\,\,\,\,\,\,\,\,\,\,x+0.49 = x^\prime\cos14^\circ-y^\prime\sin14^\circ \\
 &\,\,\,\,\,\,\,\,\,\,\,\,\,\,\,\,\,\,\,\,\,\,\,\,\,\,\,\,\,\,\,\,\,\,\,\,y+0.005 =x^\prime\sin14^\circ+y^\prime\cos14^\circ .
\end{align*}
Using the above equations and the results obtained in the preceding section, we image a disc of uniform brightness centred on the quasar, inclined by 54\dego with respect to the plane of the sky and having a radius $R_{disc}$=1.35 arcsec. The inclination of the disc simply divides the disc plane brightness by cos54\dego to give the projected brightness. The images are convolved with the beam: the central panel of Figure ~\ref{fig6} shows that the morphology of the image brightness produced by a uniform disc is confined within an approximately elliptical annular band centred on the lens/quasar region. We define two ellipses, E+ and E$-$, delimiting the outer and respectively inner edges of the uniform disc image as shown in Figure ~\ref{fig6}. The right panel of \mbox{Figure ~\ref{fig6}} displays the image brightness of the observed data. It is well contained within ellipses E+ and E$-$ and populates the middle of the band delimited by them. In order to use coordinates adapted to this morphology, we define a parameter $\lambda$ such that $\lambda$=$-$0.5 on E$-$ and +0.5 on E+, namely $\lambda$=$\pm$0.5 on E$\pm$. Precisely, a point on the sky plane having Cartesian coordinates ($x=r\cos\omega$, $y=r\sin\omega$) and polar coordinates ($r,\,\omega$) is imaged as \mbox{($\lambda$, $\omega$)} with $\lambda=[r-(r_++r_-)/2]/(r_+-r_-)$; here, $r_+$ and $r_-$ are the points of position angle $\omega$ on ellipses E+ and E$-$ respectively. In the remaining of the section we work in this new system of coordinates, ($\lambda$, $\omega$, $V_z$).

\subsection{The data cube }
We construct a data cube in ($\lambda$, $\omega$, $V_z$) coordinates to be compared with model predictions. We set its limits as $-$0.5$<$$\lambda$$<$+0.5, 0$<$$\omega$$<$360\dego and $-$340$<$$V_z$$<$333 \kmsns; we segment it in 20 $\lambda$  bins, each 0.05 wide, in 18 $\omega$ bins, each 20\dego wide and in 16 $V_z$ bins, each $\sim$42 \kms wide. We construct the new data cube starting from the reduced data presented in Section \ref{sec3}, distributed in 125$\times$125 pixels in the sky plane, each \mbox{50$\times$50 mas$^2$} in area, and 80 Doppler velocity bins, \mbox{each 8.417 \kms wide}. Each pixel is probed in 100 points randomly distributed over its area.

\begin{figure*}[!h]
  \centering
  \includegraphics[height=10.cm,trim=0.cm 0.5cm 0.cm 2.cm,clip]{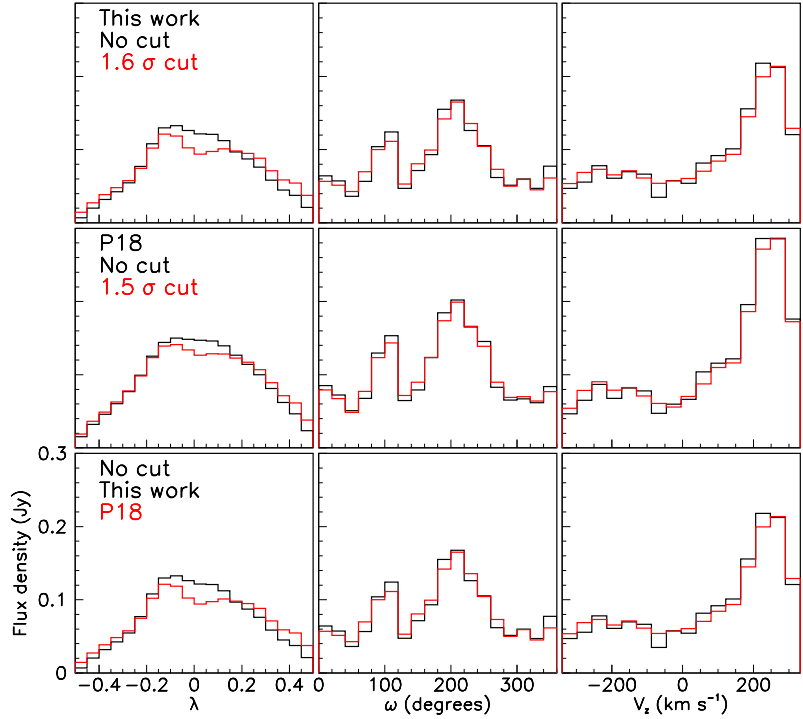}
\caption{Projections of the new data cube (see text) on $\lambda$ (left), $\omega$ (centre) and $V_z$ (right). The upper row compares the present data without cut (black) or with a 1.6-$\sigma$ cut (red) corresponding to 11.7 $\mu$Jy per original data cube element (50$\times$50$\times$8.417 mas$^2$ \kmsns). The second row compares the P18 data without cut (black) or with a 1.5-$\sigma$ cut (red) corresponding to 7.1 $\mu$Jy per original data cube element. The third row compares the present data (black) with the P18 data (red).}
 \label{fig7}
\end{figure*}

Working in the sky plane rather than in the uv plane prevents in practice a rigorous treatment of the noise. Instead, we compare in Figure ~\ref{fig7} the distributions obtained by applying no brightness cut on the original data cube elements to those obtained by applying a cut ($\sim$1.5-$\sigma$) such that the flux integrated over the whole data cube is the same ~\cite{Thai+2020}. The upper row of Figure ~\ref{fig7} compares the projections of the data cube on each of the coordinates with and without application of the cut. We repeat the exercise with the data reduced by P18. Applying a cut (again \mbox{$\sim$1.5-$\sigma$}) producing the same integrated flux as when no cut is applied gives the results shown in the second row of \mbox{Figure \ref{fig7}}. Finally, the third row compares the present data with the P18 data, both with no brightness cut being applied.

In summary, the differences in noise level and angular resolution between the analyses of the present work and of P18, apart from a global rescaling factor ($\sim$0.8) that is irrelevant to the following arguments, are unimportant. Differences between the four data cubes illustrated in \mbox{Figure ~\ref{fig7}} help with an evaluation of the uncertainties attached to these measurements, on average $\sim$13 mJy per histogram bin.

\subsection{The disc model}
\begin{figure*}[!h]
  \centering
  \includegraphics[height=4.5cm,trim=0cm 1.3cm 0.cm 2.2cm,clip]{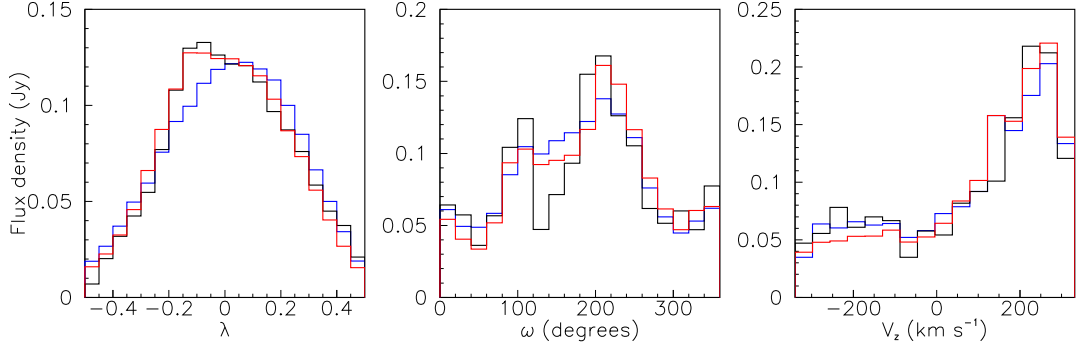}
   \caption{Comparison between observations (black) and best fit model (red and blue) are shown as projections of the data cube on $\lambda$ (left), on $\omega$ (centre) and on $V_z$ (right). The red (blue) histograms are with (without) allowance for the presence of an emission hot spot (see Section ~\ref{sec4.4}).}
  \label{fig8}
\end{figure*}

We parameterize the rotation curve as $V(R)=V_0\frac{(e^{R/R*}-1)}{(e^{R/R*}+1)}$ and the disc brightness is taken uniform over a disc of mean radius $R_{disc}$ smeared  radially by a Gaussian of dispersion $\sigma_{disc}$.This choice has been made after having explored results obtained using different possible forms. While beam convolution is properly accounted for by the model, no noise contribution is included. We decided to do so after having produced fits accounting for a Gaussian noise contribution mimicking that present in the observed data cube. The results were essentially unaffected. Optimization of the values of the four parameters is made by minimizing a $\chi^2$ function describing the quality of the match between model and observations. Rather than evaluating the value of $\chi^2$ as a sum over the 20$\times$18$\times$16=5760 data cube elements, we find that restricting the sum to the 20+18+16=54 bins of the histograms displayed in Figure ~\ref{fig7} is better adapted to the precision allowed by the quality of the data and the crudity of the model. The $\chi^2$ is evaluated using as uncertainty a common arbitrary value of 10 mJy per histogram bin and is divided by the number of degrees of freedom. The best fit is obtained for $V_0$=405 \kmsns, $R^*$=0.22 arcsec (1.6 kpc), $R_{disc}$=1.10 arcsec (7.7 kpc) and \mbox{$\sigma_{disc}$=0.32 arcsec (2.2 kpc)}. It corresponds to a steeper rise of the rotation curve at the origin than implied by P18 and L17. The results are illustrated in Figure ~\ref{fig8}.

The best fit value of $\chi^2$ is $\sim$3, meaning that the mean deviation between model and of observation is $\sim$17 mJy per histogram bin. As could be expected, the model is unable to account for the observed brightness inhomogeneity, particularly well revealed by the central panel of Figure ~\ref{fig8}, which we discuss in the next section.
\begin{figure*}[!h]
  \centering
  \includegraphics[height=4.cm,trim=-1.cm 1cm .2cm 1.cm,clip]{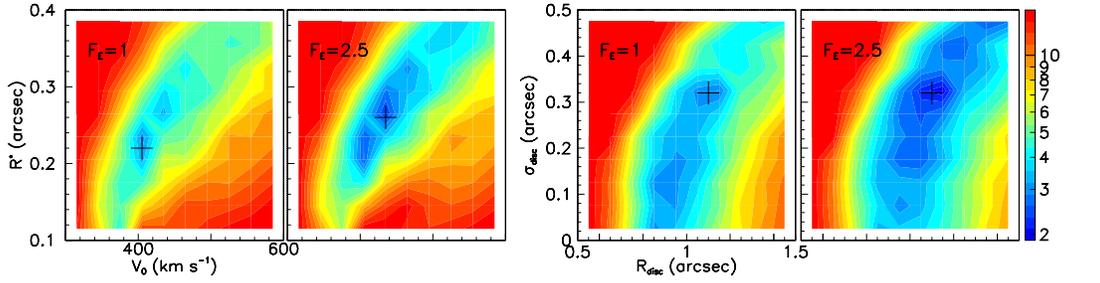}
 \caption{Left: dependence of $\chi^2$ on the model parameters: $V_0$ (abscissa) and $R^*$ (ordinate) in the left-most panels and $R_{disc}$ (abscissa) and $\sigma_{disc}$ (ordinate) in the right-most panels. In each pair of panels, the results obtained without and with allowance for an excess of emission in the vicinity of the quasar are ordered from left to right. In each panel the other model parameters are set at their best-fit values. The values of $F_E$ listed in the inserts refer to the excess of emission discussed in Section ~\ref{sec4.4}.}
  \label{fig9}
\end{figure*}
The results obtained for the rotation curve and for the disc brightness are essentially uncorrelated; but the results obtained for each independently display a very strong correlation between $V_0$ and $R^*$ on the one hand and between $R_{disc}$ and $\sigma_{disc}$ on the other. This is illustrated in Figure ~\ref{fig9} which maps the value of $\chi^2$ in the $R^*$ vs $V_0$ and $\sigma_{disc}$ vs $R_{disc}$ planes respectively.

\subsection{Brightness inhomogeneity}\label{sec4.4}
\begin{figure*}[!h]
  \centering
  \includegraphics[height=10.cm,trim=0.cm 1.cm 0cm 1.5cm,clip]{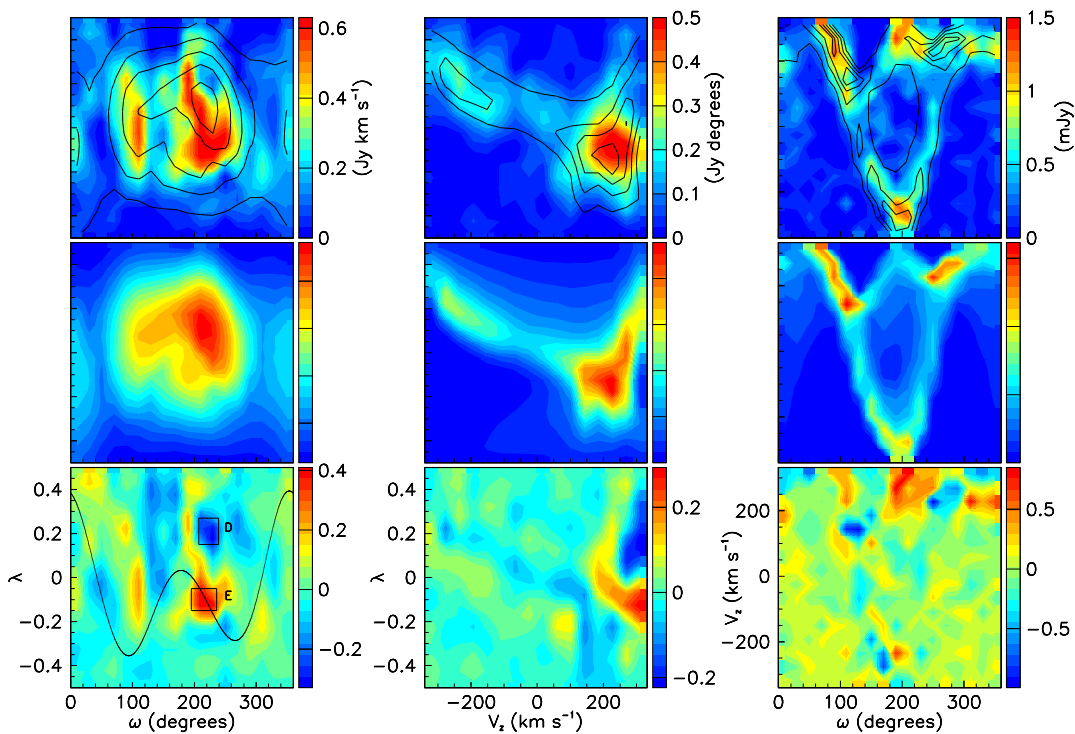}
  \caption{Maps of the brightness projected on the ($\lambda$ vs $\omega$), ($\lambda$ vs $V_z$) and ($V_z$ vs $\omega$) planes (respectively from left to right) and integrated over the third coordinate, $V_z$, $\omega$ and $\lambda$ respectively. Bins are 0.05 wide in $\lambda$, 20\dego wide in $\omega$ and 42 \kms wide in $V_z$. Upper panels display observed images with no brightness cut. Central panels display the images obtained from lensing the best fit disc model. Contours of the disc model (central row) are superimposed on observations (upper row).  Lower panels map the difference between observations and model. The rectangles in the left panel show the regions of strong inhomogeneity discussed in Section 4.4, the critical curve (black) is also plotted.}
  \label{fig10}
\end{figure*}

\begin{figure*}[!ht]
  \centering
  \includegraphics[height=4.2cm,trim=0.cm 0.9cm 2.2cm .2cm,clip]{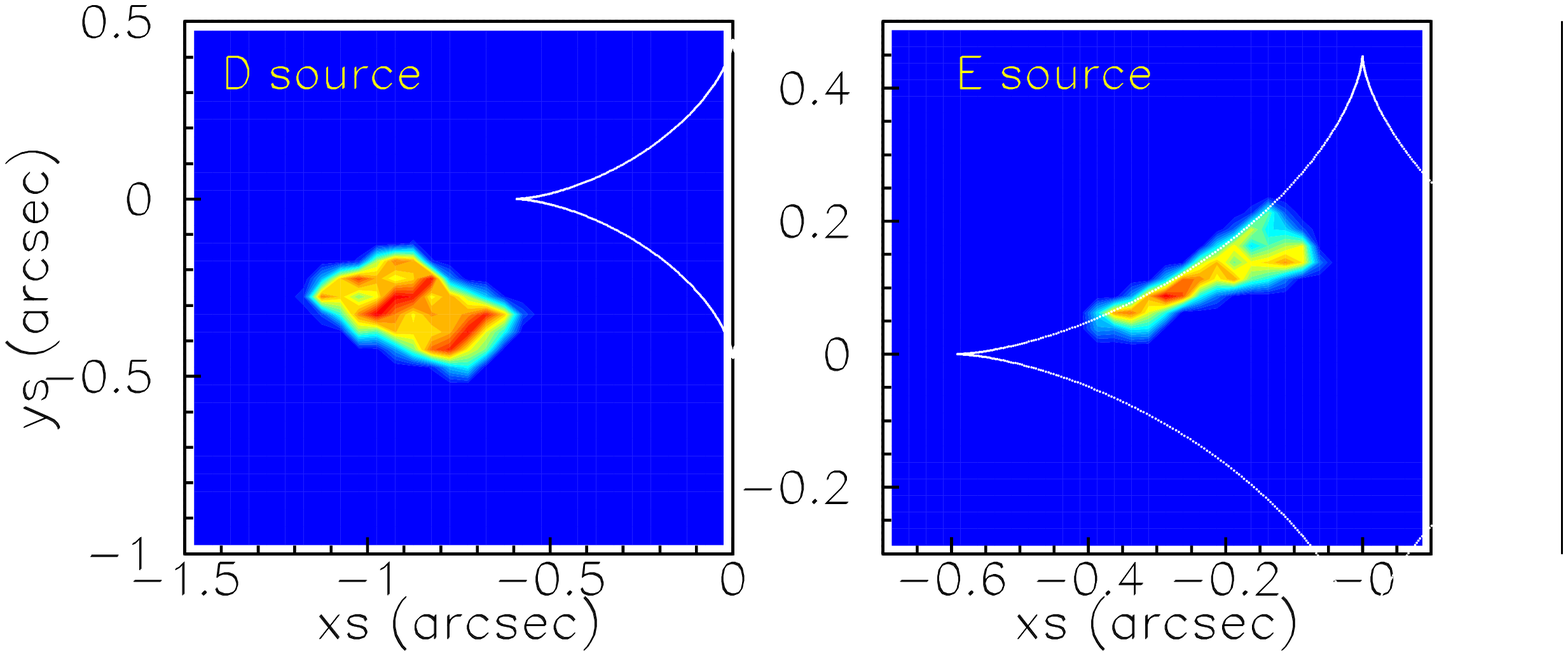}
  \includegraphics[height=4.2cm,trim=0.cm 0.9cm 3.cm .2cm,clip]{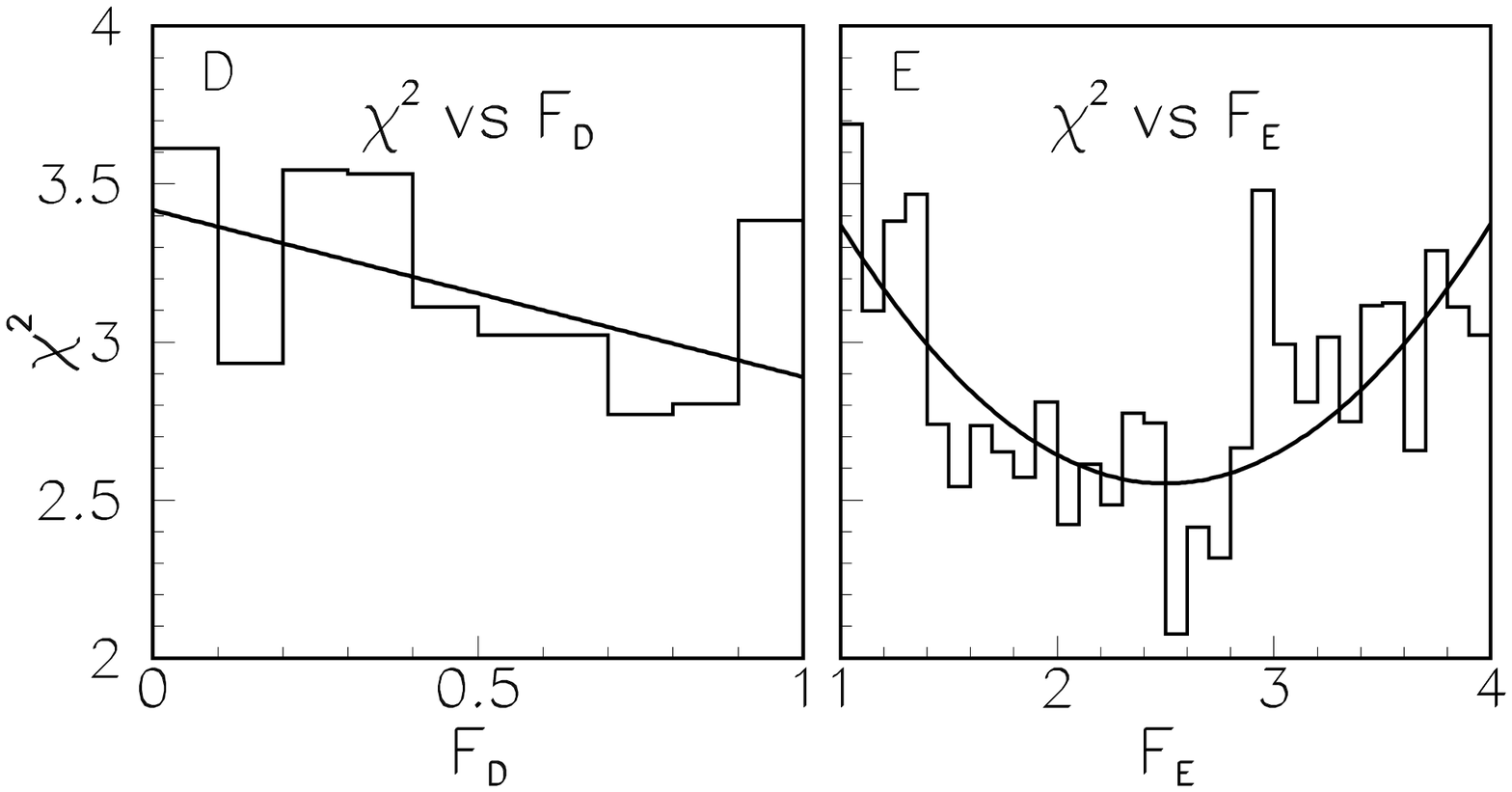}
  \caption{Left panels display the brightness distribution in the source plane obtained by de-lensing depletion D (left) and excess E (centre-left) respectively. Right panels display the dependence of $\chi^2$ on the factors $F_D$ and $F_E$ measuring the amplitudes of the depletion (centre-right) and excess (right) respectively when the other model parameters are fixed at their best-fit values.}
  \label{fig11}
\end{figure*}

Maps of the observed and modelled brightness are compared in Figure ~\ref{fig10} in the $\lambda$ vs $\omega$, $\lambda$ vs $V_z$ and $V_z$ vs $\omega$ planes. In general, the agreement between observed and modelled maps is remarkable given the crudeness of the model. The $V_z$ vs $\omega$ maps display an oscillation that reveals very clearly the disc rotation. The strong asymmetry of the Doppler velocity spectrum, present in both model and observed brightness distributions, is the direct result of the lens caustic being entirely located in the red-shifted region of the rotating disc. 	These maps provide important information on the morpho-kinematics without requiring de-lensing of the observed images.
Differences between model and observation are illustrated in the lower panels of the figure. They are dominantly in the two red-most $V_z$ bins and confined to a narrow interval of the polar angle $\omega$, approximately between 200\dego and 240\degons, where they take the form of an excess for small negative values of $\lambda$ and a depletion for small positive values. We define two rectangles in the $\lambda$ vs $\omega$ plane that delimit these regions: an excess E (200\degons$<$$\omega$$<$235\degons,$-$0.18$<$$\lambda$$<$$-$0.05) and a depletion D (207\degons$<$$\omega$$<$237\degons,0.15$<$$\lambda$$<$0.27). The left panels of Figure ~\ref{fig11} show the associated regions of the source plane.

The source of D is outside the caustic and its magnification spans values between $\sim$3 and $\sim$5. The companion image of D is small and centred at ($x$, $y$)$\sim$(0.75, 0.25) \mbox{arcsec}; its magnification spans values between 0.5 and 1. In contrast, the source of E is located inside the caustic and touches its edge; its magnification spans accordingly a very broad range of values starting at $\sim$4 and extending to infinity with a mean value of $\sim$20. It produces three additional images, which span magnifications ranging between 1.5 and 5.5; one of these, centred at ($x$, $y$)$\sim$($-$1, 2) arcsec is in a region where the lower-left panel of Figure ~\ref{fig10} shows that the observed brightness exceeds slightly that of the model. The fact that de-lensing E produces a source limited by the caustic curve is not an artefact of the lensing mechanism. In fact, the source that produces E overlaps the caustic but only the part of it that is inside contributes to image E. The part that is outside produces instead additional images, one located near ($x$, $y$)$\sim$($-$1, 2) and the other close to E.

Interpreting the inhomogeneity as the conjunction of an excess and a depletion is arbitrary; it assumes that on average observed and modelled fluxes are equal, which they have no reason to be; it is indeed simpler and more natural to blame the inhomogeneity on a single hot spot, in which case the flux predicted by the model needs to be scaled down, or as a single depletion, in which case the flux predicted by the model needs to be scaled up. We tried both interpretations by including in the model an enhancement (depletion) of emission by a factor $F_E$ ($F_D$) in the source region of E (D) shown in Figure ~\ref{fig11}. In both cases the best fit values of the model parameters, $V_0$, $R^*$, $R_{disc}$ and $\sigma_{disc}$, were essentially unaffected but allowing for a depletion did not produce lower values of $\chi^2$; in contrast, as illustrated in the right panels of Figure ~\ref{fig11}, allowing for an excess resulted in a significant improvement of the quality of the fit. The best fit values of the model parameters are now $V_0$=435 \kmsns, $R^*$=0.26 \mbox{arcsec} (1.8 kpc), $R_{disc}$=1.10 arcsec (7.7 kpc), $\sigma_{disc}$=0.32 arcsec (2.25 kpc) and $F_E$=2.5 (see Figure ~\ref{fig9}).

The above results call for a word of caution. Qualitatively, the presence of an excess of emission in the vicinity of the quasar is likely to be a robust result of the analysis; it had been noted by P18 who suggest that it is most likely associated with a site of on-going star-formation. However, its location on top of the caustic makes it difficult to specify reliably its morpho-kinematical properties. The lens model used in the present work is very crude and so is also that used by P18: none of these allows for local inhomogeneity of the lensing potential, they describe it in very general terms; the vicinity of the caustic is particularly difficult to model reliably, small variations of the lensing potential having strongly amplified effects in the construction of the images. As we shall see in the next section, the overall remarkable agreement between model and observations displayed in \mbox{Figure ~\ref{fig10}} does not sustain a significantly more detailed scrutiny.
 
\subsection{Rotation and turbulence}
\begin{figure*}[!ht]
  \centering
  \includegraphics[height=5cm,trim=0.cm 1.5cm 1.cm 1.cm,clip]{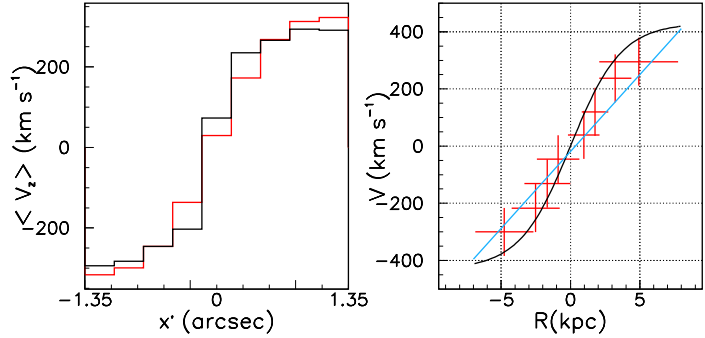}
 \caption{Dependence of $<$$V_z$$>$ on $x^\prime$ (major axis of the projection of the disc on the plane of the sky) for the data (black) and the model (red) is shown in the left panel and the rotation curve in the right panel together with the L17 (blue curve) and P18 (red crosses) results.}
 \label{fig12}
\end{figure*}

The evaluation of the rotation curve made by P18 and L17 consists in defining a band bracketing the major axis of the projection of the disc on the plane of the sky ($x^\prime$ axis in Figure ~\ref{fig6}).  We choose for it a width of $\pm$1 kpc and a length of 2.7 arcsec ($\sim$19 kpc), divided in nine segments, each having a length of 0.3 arcsec ($\sim$2 kpc). In each of these segments we compare the observed and modelled velocity spectra. The modelled spectra are obtained by de-lensing the images produced by lensing the model disc source and convolved with the beam. Results are illustrated in the left panel of \mbox{Figure ~\ref{fig12}}. Qualitatively, the general trend is well reproduced by the model but significant differences are observed in the central segments: the data display larger Doppler velocities on the red side and lower Doppler velocities on the blue side than implied by the model. Moreover, in the central segment, the line width predicted by the model is much smaller than that observed in the data ~\cite{Thai+2020}. A natural interpretation of such an effect is disc warping causing an effective dependence on $\theta$ of the sine of the inclination angle in Relation (\ref{rela4}). However, including warping in the model by writing $V_z=V(R)\cos\theta\sin\varphi$ with $\varphi$ depending simply on $\theta$ and $R$, gives only a modest improvement of the match between model and observations. This suggests that a more complex dynamics than described by the simple model is at stake.   

Beam convolution causes an effective smearing of the disc region contributing to each of the nine segments, making the radial distribution of the mean Doppler velocity measured in the central segments less steep and causing an increase of the velocity dispersion. This important result is a warning: measuring the velocity dispersion requires a careful evaluation of the contribution of rotation within the finite angular resolution of relevance. Making the angular acceptance smaller will decrease the direct contribution of rotation, independent from beam convolution, but will not decrease the contribution resulting from the smearing caused by the beam. This contribution is important: using the central segment as an example, the line width predicted by the model increases from 60 \kms to nearly 100 \kms when accounting for beam convolution, namely a beam contribution of over 70 \kms (both contributions add up in quadrature).

The differences between observed and predicted velocity dispersions may receive contributions from the intrinsic line width (turbulence), which is absent from the model, or from different rotation contributions, as could be caused by disc warping. We recall that P18 claim that the observation of a high dispersion in the vicinity of the quasar demonstrates that the region of enhanced emission studied in the preceding section is the seat of increased gas turbulence. However, the differences observed in the central segments of the band used to evaluate the rotation curve do not consist of a simple broadening of the line, as would be expected from turbulence, but reveal rather an excess of red-shifted emission.

Maps of the intensity, mean Doppler velocity and Doppler velocity dispersion in the $\lambda$ vs $\omega$ plane reveal important differences between model and observations ~\cite{Thai+2020}: predicted mean Doppler velocities tend to be too small in the 90\degons$<$$\theta $$<$180\dego quadrant and too large in the \mbox{270\degons$<$$\theta$$<$360\dego} quadrant of the disc plane; velocity dispersions reach some 120 \kms for the model and some \mbox{160 \kms} for the observations. We failed to obtain a substantial improvement of the quality of the match between model and observations by including a simple description of disc warping in the model: this confirms the need for a more complex dynamics than implied by the simple disc model.  From a number of comparisons between the line widths predicted by the model with and without beam convolution, the contribution is found to be in the range between 50 and 70 \kmsns.

\begin{figure*}
  \centering
  \includegraphics[width=.8\textwidth,trim=.1cm 1.2cm 4.cm 1.cm,clip]{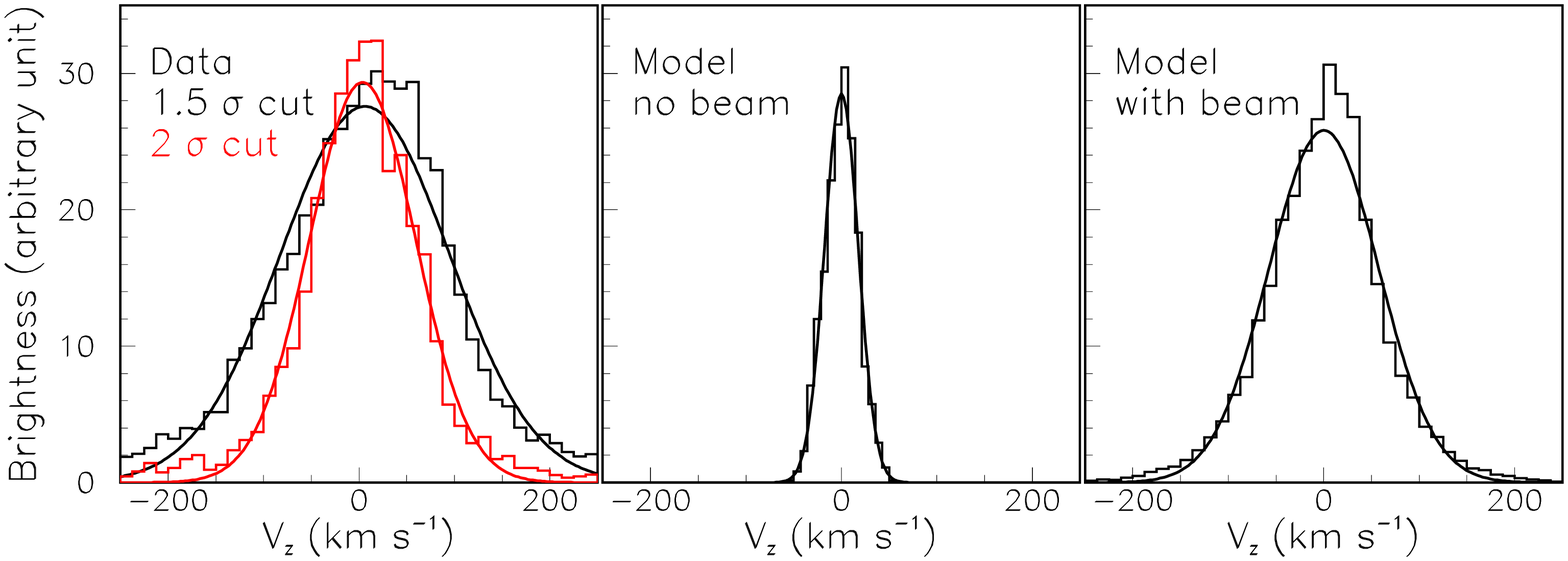}
  \caption{Line profiles in the region $|\lambda|<$0.25 and 110\degons$<$$\omega$$<$250$^\circ$. Left: observations with cuts at 11.7 $\mu$Jy (black) or 16 $\mu$Jy  (red); the lines show Gaussian fits with $\sigma$=85 \kms and 56 \kms respectively. Gaussian fits to the predictions of the disc model without (centre) and with (right) beam convolution give $\sigma$ values of 18 \kms and 58 \kms respectively.}
  \label{fig13}
\end{figure*}

An independent evaluation is obtained by selecting the central region of the $\lambda$ vs $\omega$ plane (left column of \mbox{Figure ~\ref{fig10}}) defined as $|\lambda|<$0.25 and $|\omega-180^\circ|<$70$^\circ$. We segment it in 10$\times$14 pixels of 0.05 in $\lambda$ and 10\dego in $\omega$. For each pixel ($i,j$) we evaluate the mean value $<$$V_z$$>_{ij}$ of the Doppler velocity and show the distribution of $V_z-$$<$$V_z$$>$$_{ij}$ in Figure ~\ref{fig13}. We obtain this way an evaluation of the line profile. The observed line widths ($\sigma$) are 85 and 56 \kms for cuts at 11.7 (1.5-$\sigma$) and 16 $\mu$Jy (\mbox{2-$\sigma$) per 50$\times$50$\times$8.417 mas$^2$ \kms respectively}.  As the model includes no intrinsic line broadening, the line broadening shown by the model is entirely due to the contribution of rotation. Gaussian fits give $\sigma$ values of 18 and \mbox{58 \kms} without and with beam convolution respectively, implying a contribution of \mbox{$\sim$50 \kms} from the smearing effect of the beam size. The line width measured with the 16 $\mu$Jy cut does not require any contribution of turbulence while that measured with an 11.7 $\mu$Jy cut allows for a contribution reaching $\sim$60 \kmsns: without a precise understanding of the contribution of noise, it is therefore difficult to obtain a reliable evaluation of the line broadening caused by turbulence.

In summary, a simple rotating disc model gives a global picture that describes well the general trend of the observed kinematics. However, such a model fails to reproduce quantitatively the details of the Doppler velocity distribution: it reveals a more complex dynamics. Attempts to describe the mismatch as the result of a simple disc warping have failed. An important result of our analysis is that the smearing in the image plane caused by the beam size contributes between 50 and 70 \kms to the dispersion of the Doppler velocity. It combines in quadrature with the contribution of rotation within the angular acceptance being probed by the pixel size of relevance. Taken together, these effects make it difficult to evaluate reliably the contribution of turbulence to the line width. But they prevent claiming that such contribution is important: within experimental uncertainties the line width can be accounted for by the dispersion of rotation velocities within the beam size.
\section{Summary and conclusions} \label{sec5}
The present study of the emission of the CO(2-1) molecular line by the host galaxy of quasar RX J1131 uses ALMA observations of unprecedented quality; they have been previously analysed in much detail by their proponents, ~\cite{Paraficz+etal+2018}, P18. We have shown that the HST images of the quasar and of the lens galaxy are very well reproduced by a simple lensing potential, sphere+external shear. We obtained parameters that are in agreement with the results of previous authors within errors. We discussed the uncertainties attached to the model parameters and their correlations and explained them by remarking that these results probe only the environment of the eastern cusp of the caustic. At this stage, the validity of the model at larger distances, in the region covered by the emission of the host galaxy, cannot be taken as granted. But we demonstrated its validity from our analysis of the ALMA observations of the CO(2-1) line emission. We reduced the raw data and produced cleaned images, at variance with P18 who perform their analysis in the uv plane. We obtained morpho-kinematics properties of the gas that are in good agreement with the results of P18. We used two different methods for de-lensing these images, which produce source brightness distributions in agreement with each other as well as with the P18 results. We have shown that the general morphology of the image is dictated by the properties of the lens and confined within a band bracketing the critical curve. This leaves in practice no freedom to conceive lensing potentials that would be significantly different from those that we and other authors have been using. It guarantees the robustness of the results obtained by P18 within observational uncertainties.

We used polar coordinates better adapted to this geometry than Cartesians to reveal very clearly the evidence for rotation. We looked for possible deviations from the predictions of a simple rotating disc model, in particular for the possible presence of a companion as was predicted by L17 using Plateau de Bure observations of much less good angular resolution than the ALMA observations; but our analysis did not support such a presence. We paid special attention to the red-most velocity interval, which displays a particularly complex morphology. We found evidence for enhanced emission revealing a significant deviation from the prediction of the simple disc model. However the associated hot spot in the source plane overlaps the caustic, implying important uncertainties on its de-lensed properties.

We discussed the rotation curve in some detail and argued that it probably rises faster than assumed by P18 in the vicinity of the quasar; we noted that the angular resolution is such that the dispersion of the rotation velocity within the beam size in the image plane causes an important effective broadening of the line width, which we evaluated in the range between 50 and 70 \kms. This is sufficient to account for the observed line width when applying a $\sim$2-$\sigma$ cut on the data. However, uncertainties attached to the effect of noise prevent from giving a reliable evaluation of the contribution of turbulence. This is at variance with the conclusion of P18 who attribute the high Doppler velocity dispersion observed in the vicinity of the AGN to \mbox{gas turbulence exclusively}.

P18 obtained from their analysis a number of results concerning the physical properties of the galaxy, which they compared with those of other similar galaxies. The nature of our study prevents us from adding much to their conclusions. We simply remark that their estimate of the total dynamical mass enclosed within 5 kpc, (1.46$\pm$0.31)$\times$10$^{11}$ \Msun, may be affected by the steepness of the rotation curve: at fixed radius, the dynamical mass is proportional to the square of the rotation velocity. The analysis presented in the preceding sections implies therefore a possibly higher value of the gas mass, obtained by P18 and accordingly affect the value of the CO$-$H$_2$ conversion factor.

\section*{ACKNOWLEDGMENT}
 We express our deep gratitude to Professors Frederic Courbin and Matus Rybak who kindly provided us with copies of data files summarizing the results of the P18 analysis. This paper uses archival ALMA data from project 2013.1.01207.S (PI: Paraficz Danuta). ALMA is a partnership of ESO (representing its member states), NSF (USA), NINS (Japan), NRC(Canada), NSC/ASIAA (Taiwan), and KASI (South Korea), in cooperation with Chile. The Joint ALMA Observatory is operated by ESO, AUI/NRAO and NAOJ. We are deeply indebted to the ALMA partnership, whose open access policy means invaluable support and encouragement for Vietnamese astrophysics. Financial support from the World Laboratory and VNSC is gratefully acknowledged. This research is funded by the Vietnam National Foundation for Science and Technology Development (NAFOSTED) under grant number 103.99-2019.368.


\bibliographystyle{cip-sty-2019}
\bibliography{ref}

\end{document}